\documentclass[a4paper, 11pt]{article}
\usepackage{graphicx}
\usepackage{moriond}
\newcommand{\uu}[1]{\, \mathrm{#1}} % for /u/nits in math mode

\begin{document}
\vspace*{4cm}
\title{THE CRESST-II EXPERIMENT}
\author{RAFAEL~F.~LANG}
\address{Max-Planck-Institut f\"ur Physik, F\"ohringer Ring 6, D-80805 M\"unchen, Germany, \\
E-mail rafael@cresst.de}
\author{For the CRESST Collaboration: G.~Angloher, M.~Bauer, I.~Bavykina, A.~Bento, A.~Brown, C.~Bucci, C.~Ciemniak, C.~Coppi, G.~Deuter, F.~von~Feilitzsch, D.~Hauff, S.~Henry, P.~Huff, J.~Imber, S.~Ingleby, C.~Isaila, J.~Jochum, M.~Kiefer, M.~Kimmerle, H.~Kraus, J.~Lanfranchi, R.~F.~Lang, M.~Malek, R.~McGowan, V.~B.~Mikhailik, E.~Pantic, F.~Petricca, S.~Pfister, W.~Potzel, F.~Pr\"obst, S.~Roth, K.~Rottler, C.~Sailer, K.~Sch\"affner, J.~Schmaler, S.~Scholl, W.~Seidel, M.~Stark, L.~Stodolsky, A.~J.~B.~Tolhurst, I.~Usherov, W.~Westphal.}
\maketitle
\abstracts{
The CRESST-II experiment in the Laboratori Nazionali del Gran Sasso uses scintillating crystals as a target to search for elastic scatterings of dark matter particles in a laboratory environment. The detectors are operated in a dilution cryostat at temperatures below $30\uu{mK}$, and for each particle interaction, the phonon signal as well as the scintillation light signal are recorded. The current limit that can be placed on the spin-independent WIMP-nucleon scattering cross-section with this technique is below $6\times10^{-7}\uu{pb}$ for WIMPs in the mass range from about $40$ to $90\uu{GeV/c^2}$.}

\section{Introduction}
It remains one of the most pressing problems of physics today to clarify the nature of dark matter. So far, the evidence for a significant component of non-baryonic dark matter in the universe is compelling. It stems from a variety of independent experiments, ranging from measurements of the cosmic microwave background~\cite{hinshaw2008} and big bang nucleosynthesis~\cite{steigman2007} to astrophysical observations of large amounts of unseen matter in galaxy clusters~\cite{clowe2007} and galaxies~\cite{xue2008}. Yet, the nature of this component is still unknown, so experiments to unravel this mystery are more important than ever.

Highly motivated candidates for dark matter are weakly interacting, yet massive particles (WIMPs)~\cite{jungman1996}. These particles are expected to be gravitationally bound in the Milky Way in a roughly isothermal halo, thus following a Maxwell-Boltzmann-distribution~\cite{donato1998,lewin1996}. The WIMP density at the position of the Earth is assumed to be around $0.3\uu{GeV/cm^3}$. On our way around the Galaxy we pass through this halo with a velocity of about $220\uu{km/s}$, and we hope to detect the scattering of this WIMP wind on nuclei in an absorber.

Accelerator searches for new forms of matter place a lower bound on the mass of these particles of the order of $10\uu{GeV/c^2}$. Since the de~Broglie wavelength of particles with such mass and velocity is larger than the radius of a nucleus, one may expect the scattering to occur in a coherent way, the case we consider here. Then, for a given target material with mass number $A$, the scattering amplitude scales as $A^2$, favoring heavy nuclei for a search experiment.

\section{Scintillating Crystals}

Following the interaction of a WIMP with a target nucleus, the energy of the recoil is very small, typically only of the order of $10\uu{keV}$. Together with the very small expected interaction rate of less than 10 events per kilogram of target and year of measuring time, available technologies are highly constrained. So the two main requirements to this kind of experiment are a very low energy threshold and the capability to reject backgrounds caused by known particle species.

\begin{figure}[hbp]\centering
\begin{minipage}{0.59\textwidth}\includegraphics[width=1.0\textwidth]{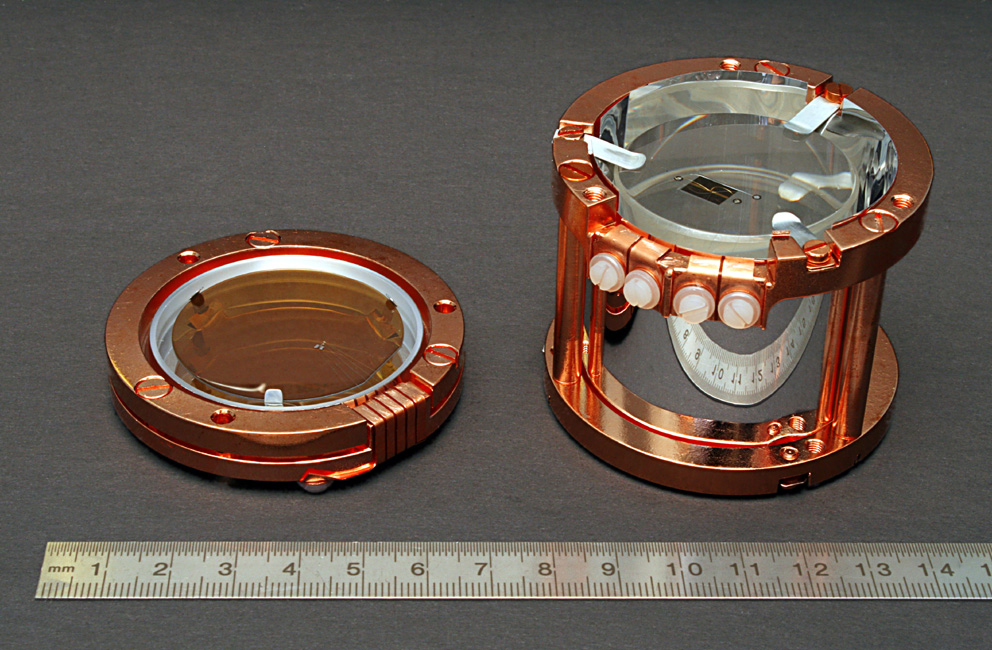}\end{minipage}
\begin{minipage}{0.39\textwidth}\includegraphics[width=1.0\textwidth]{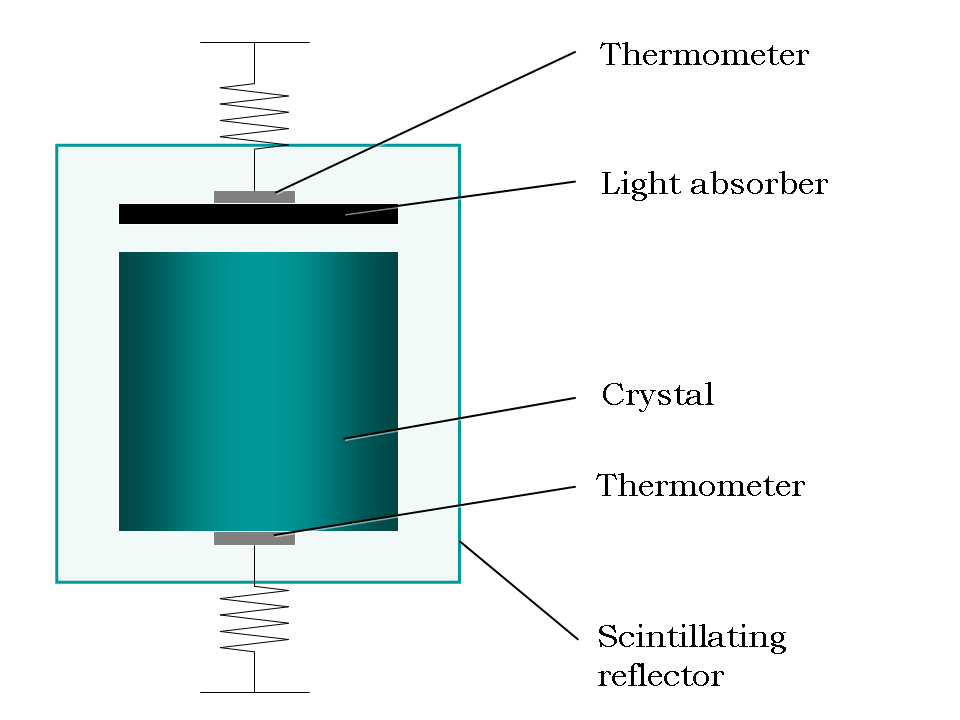}\end{minipage}
\caption{A detector module as used in the CRESST-II experiment. Left picture: An open module with the light absorbing wafer on the left hand side and the scintillating crystal in its housing on the right hand side. Sketch: a tungsten superconducting phase transition thermometer is evaporated onto a $\mathrm{CaWO_4}$ target crystal. A second thermometer is evaporated on a light absorber to measure the scintillation light. Both detectors are enclosed in a scintillating reflective housing.}\label{fig:module}
\end{figure}

In the second phase of the Cryogenic Rare Event Search with Superconducting Thermometers CRESST-II~\cite{angloher2002} we use scintillating crystals as target material~\cite{angloher2005}. They are shaped as cylinders of $(4\times4)\uu{cm}$ and weigh about $300\uu{g}$ each. We operate them as calorimeters at temperatures as low as $10\uu{mK}$. In such dielectric crystals, most of the energy of a particle interaction goes into phonons. To measure these phonons, we evaporate a thin tungsten film onto the crystal. Tungsten becomes superconducting in this temperature range, and stabilizing the film in its superconducting phase transition (by means of a dedicated heater) makes an extremely sensitive thermometer: Any particle interaction warms up the film, thus changing its resistance, which can be measured with sensitive electronics~\cite{henry2007}.

In addition to this phonon signal, a small fraction of the interaction energy (typically a few percent) is emitted as scintillation light. To detect this light, we use another tungsten phase transition thermometer on a separate light absorbing wafer. Thus, for each target crystal, we have two thermometers, one measuring the deposited energy, and the other measuring the scintillation light, see figure~\ref{fig:module}.

We define the \textit{light yield} of an interaction as the amount of energy in the light detector divided by the energy in the phonon detector, and normalize it such that electron recoils have a light yield of~1. Electron recoils are caused by electrons and gammas that impinge onto the crystal. Compared to such events, the light yield of alpha particles is reduced by a factor of 5. Neutrons are mainly seen when they scatter from oxygen due to the kinematics of the interaction, with a light yield reduced by a factor of 10 relative to that of the electron recoils. Coherent scatterings of WIMPs are expected to take place mainly on tungsten where the light yield is reduced by a factor of 40~\cite{ninkovic2006}. Thus, simultaneously measuring the light signal allows us to discriminate the (possibly WIMP induced) tungsten recoils from the dominant radioactive backgrounds.

\section{The Upgraded Setup}

In 2007, the experimental setup was extended to be capable of housing up to 33 detector modules. To shield them as much as possible from ambient radioactivity, we provide a variety of shielding layers. The experiment is hosted in the Laboratori Nazionali del Gran Sasso, Italy, to shield the detectors from cosmic ray induced backgrounds with the overburden of $1400\uu{m}$ of rock. During the upgrade, we added a muon veto to discriminate residual muon induced backgrounds, as well as a $45\uu{cm}$ thick wall of polyethylene to moderate neutrons to energies below detection threshold. The setup is constantly flushed with nitrogen vapor, in particular to keep radon contaminated air away from the inner parts. A $20\uu{cm}$ thick lead shield and a $14\uu{cm}$ copper shield absorb gamma radiation coming from outside. Also, the fivefold thermal shielding of the cryostat provides an additional $1.2\uu{cm}$ thick copper shield. All the materials in the vicinity of the detectors are selected for radiopurity and handled in a clean room environment.

\begin{figure}[htbp]\centering
\includegraphics[width=.45\textwidth]{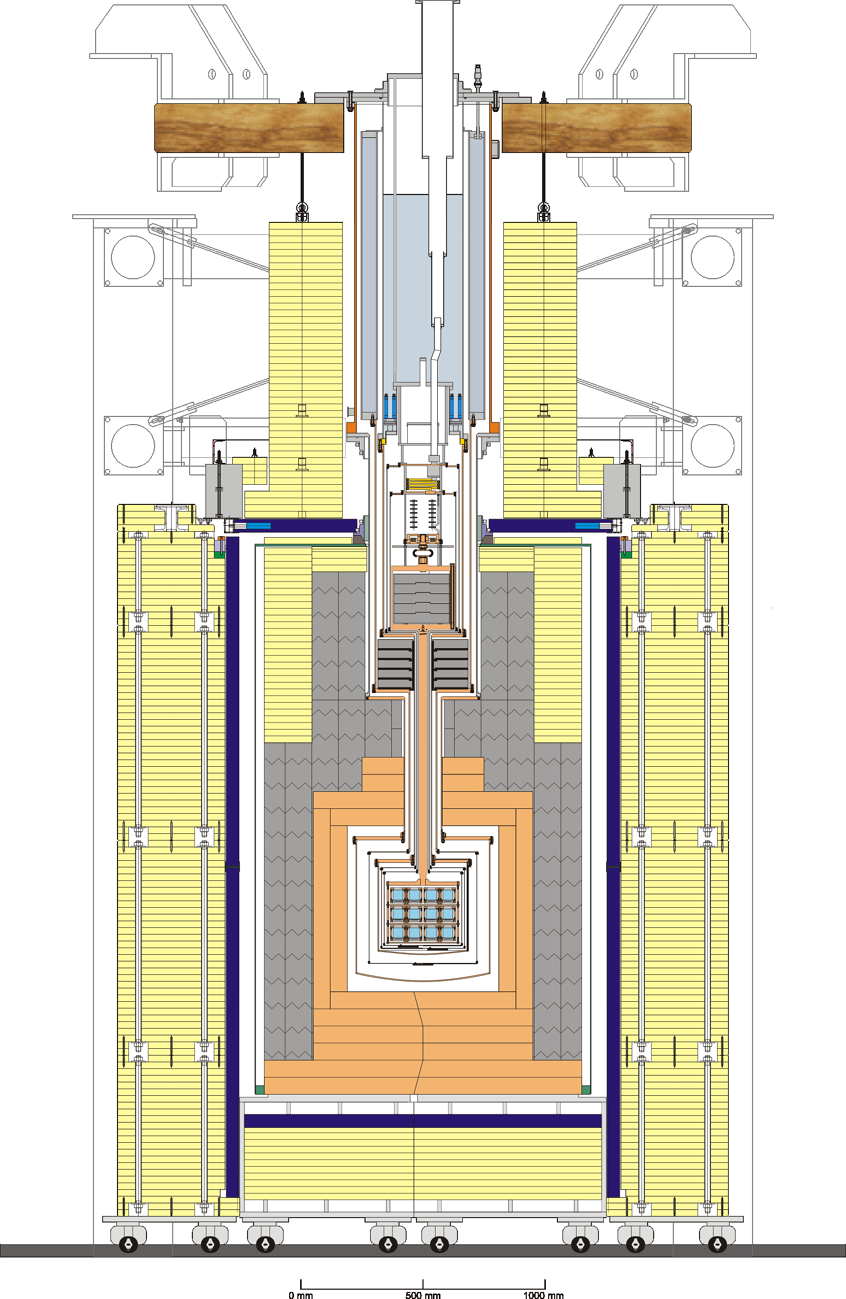}
\caption{Setup of CRESST-II. The detectors are in a low-background environment in the center of the shielding. The cryostat (upper part of the figure) is kept away from this very clean environment. The polyethylene neutron shield is shown in yellow, the lead shield in gray, and the copper shield in orange. A muon veto made from plastic scintillator, shown in blue, helps to further reduce backgrounds induced by cosmic rays.}\label{fig:setup}
\end{figure}

\section{Results}

After a major upgrade~\cite{angloher2008} we installed 10 detector modules and cooled the cryostat for commissioning of the new setup. The detectors were calibrated with a gamma source to set the energy scale, as well as a neutron source for validation of the light yield anticipated for nuclear recoils. The data reported here was taken in 2007 with two detector modules from March 27th through July 23rd.

We fit the pulses using a template fit, which is constructed from pulses from the gamma calibrations. In order to treat noise in an unbiased way, we allow the amplitude of the fit to take negative values. This guarantees that for no signal, the reconstructed amplitudes indeed scatter around zero with a width corresponding to the noise. Thus negative amplitudes can arise, resulting in events with negative light yields.

\begin{figure}[hbp]\centering
\includegraphics[width=.6\textwidth]{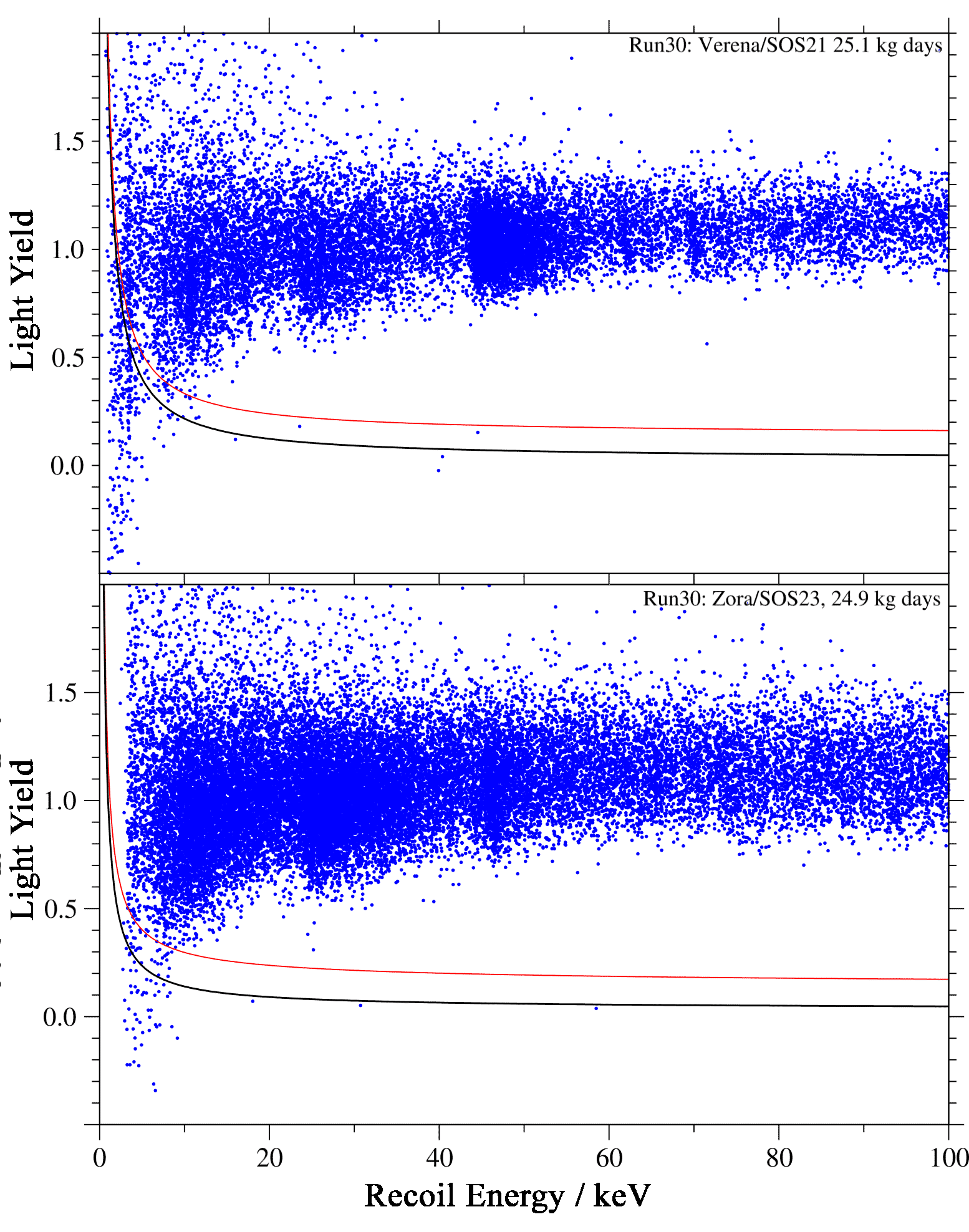}
\caption{Low energy event distribution measured with two $300\uu{g}$ $\mathrm{CaWO_4}$ detector modules during the commissioning run in 2007. The vertical axis represents the light yield expressed as the ratio (energy from the light channel)/(energy from the phonon channel), the horizontal axis is the total energy measured by the phonon channel. Below the red curve we expect 90\% of all nuclear recoils, and below the black curve 90\% of the tungsten recoils. The acceptance window is set below the black line and between $11\uu{keV}$ (above which we can discriminate electron recoils) and $40\uu{keV}$ (below which most of the WIMP induced recoils appear).} \label{fig:background}
\end{figure}

We perform only very basic quality cuts on the data, rejecting pulses only if they occur in one of the rare periods when the temperature of the cryostat is not as stable as desired, if they are direct hits of the light detector or the thermometer on the crystal, or if they are pile-up events. Events recorded from the two detectors with a cumulative exposure of $50\uu{kg\,d}$ are shown as scatter plot in figure~\ref{fig:background}.

\section{Discussion}

\begin{figure}[hbp]\centering
\includegraphics[width=.8\textwidth]{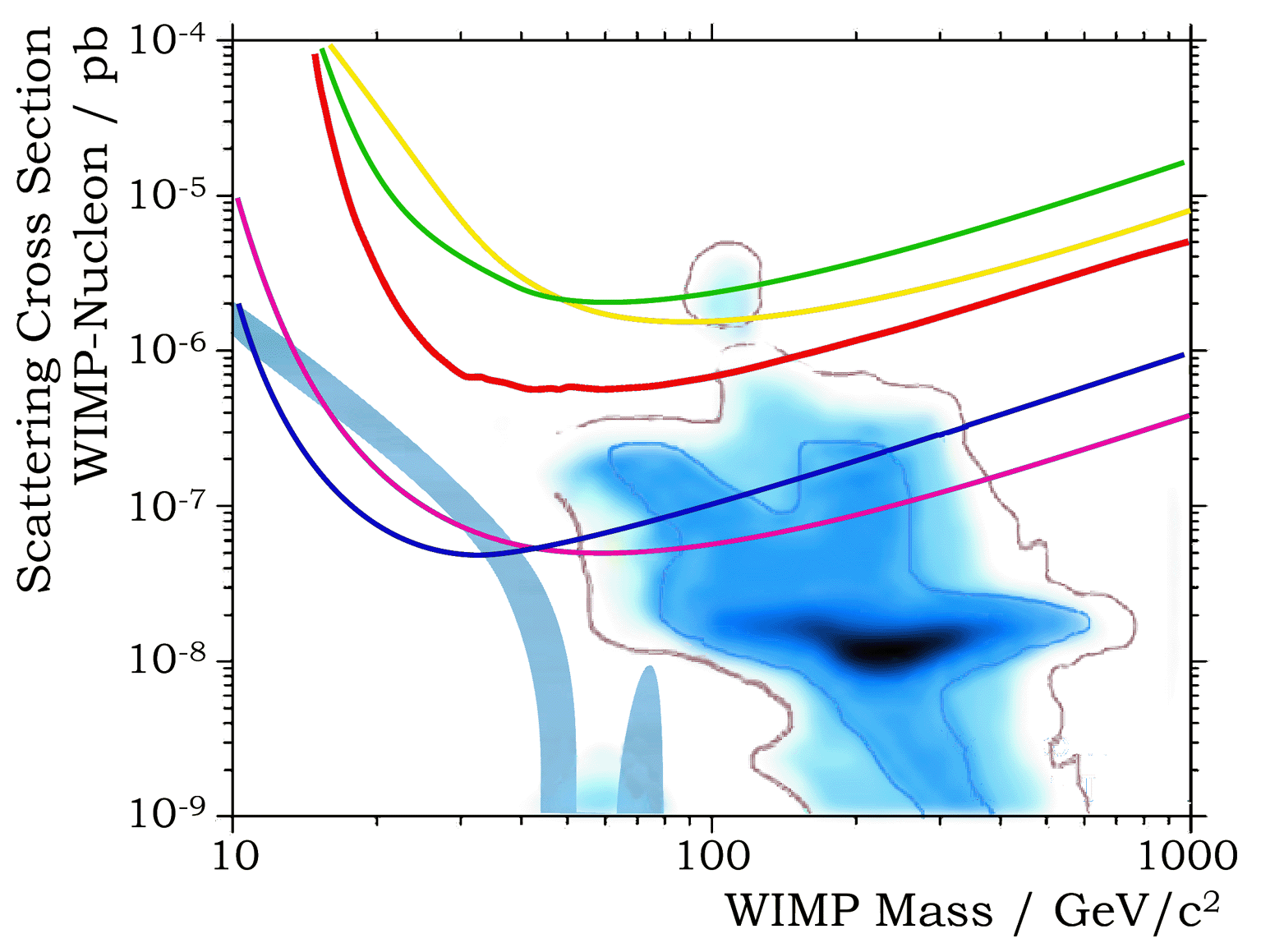}
\caption{Exclusion limits on the WIMP-nucleon spin independent scattering cross section from a few experiments, from top to bottom from the KIMS experiment (green), EDELWEISS (yellow), XENON10 (blue) and CDMS (violet). The red curve is the limit from the CRESST experiment, derived from the data of figure~\ref{fig:background} and that of the previous run. In cyan two theoretical expectations.}
\label{fig:limit}
\end{figure}

To derive a limit on the coherent WIMP-nucleus scattering cross section, standard assumptions on the dark matter halo~\cite{donato1998,lewin1996} are adopted. The finite extension of the nuclei is taken into account by assuming the Helm form factor~\cite{helm1956}, which basically limits the energy transfer to the tungsten nuclei to energies below $40\uu{keV}$ for all WIMP masses. In the energy region above $11\uu{keV}$, where recoil discrimination becomes efficient, to up to $40\uu{keV}$, 4~tungsten recoil events were observed in the data of figure~\ref{fig:background}. Combining this data with data from the previous run~\cite{angloher2005}, the upper limit for the WIMP scattering cross-section per nucleon is set using Yellin's optimum interval method~\cite{yellin2002}, shown as the red curve in figure~\ref{fig:limit}. The minimum of this curve is below $6\times10^{-7}\uu{pb}$ for WIMPs with masses between $40$ and $90\uu{GeV/c^2}$, obtained after a gross exposure (including down times due to refilling of cryogenic liquids etc.) of only $67\uu{kg\,d}$.

The few events that we observe in our signal area need not be WIMP induced tungsten recoil events. During the commissioning, a weak point in the neutron shielding above the muon veto was identified, but patched only after data taking was completed. This background can be estimated to account for the observed number of events in the nucleon recoil band. In addition we might have a small neutron background induced by decays from radioactive contaminations in the other crystals that were not running during this data taking period, a situation unique to this commissioning phase. The upcoming run should clarify on these points.

In the hunt for a discovery one might want to compare the different technologies, and since exclusion limits are one way of doing so, figure~\ref{fig:limit} also shows the results from a few other experiments: From the KIMS experiment~\cite{kims} using a gross exposure of $3409\uu{kg\,d}$ on CsI(Tl), the EDELWEISS experiment~\cite{edelweiss} ($\approx180\uu{kg\,d}$, Ge), XENON10~\cite{angle2007} ($1980\uu{kg\,d}$, Xe) and CDMS~\cite{ahmed2008} ($\approx1250\uu{kg\,d}$, Ge). The figure also contains two expectations from theory~\cite{burgess2001,roszkowski2007}.


\section*{References}
\begin{thebibliography}{88}
\bibitem{hinshaw2008} G.~Hinshaw et~al., ApJ. Supp. \textit{to be published} (2008), \texttt{arXiv:0803.0732}.
\bibitem{steigman2007} G.~Steigman, Ann. Rev. Nuc. Part. Sci. \textbf{57} (2007) 463--491, \texttt{arXiv:0712.1100}.
\bibitem{clowe2007} D.~Clowe et~al., ApJL. \textbf{648} (2007) L109--L113, \texttt{arXiv:astro-ph/0608407}.
\bibitem{xue2008} X.-X.~Xue et~al., ApJ. \textit{to be published} (2008), \texttt{arXiv:0801.1232}.
\bibitem{jungman1996} G.~Jungman et~al., Phys. Rep. \textbf{267} (1996) 195--373, \texttt{arXiv:hep-ph/9506380}.
\bibitem{donato1998} F.~Donato, N.~Fornengo and S.~Scopel, Astropart. Phys. \textbf{9} (1998) 247--260, \texttt{arXiv:hep-ph/9803295}.
\bibitem{lewin1996} J.~D.~Lewin and P.~F.~Smith, Astropart. Phys. \textbf{6} (1996) 87--112.
\bibitem{angloher2002} G.~Angloher et al., Astropart. Phys. \textbf{18}, 43--55 (2002). 
\bibitem{angloher2005} G.~Angloher et al., Astropart. Phys. \textbf{23}, 325--339 (2005).
\bibitem{henry2007} S.~Henry et al., J. Inst. \textbf{2} (2007) P11003.
\bibitem{ninkovic2006} J.~Ninkovic et al., NIM A \textbf{564} (2006) 567--578, \texttt{arXiv:astro-ph/0604094}.
\bibitem{angloher2008} G.~Angloher et al., \textit{to be published} (2008).
\bibitem{helm1956} R.~H.~Helm, Phys. Rev. \textbf{104} (1956) 1466--1475.
\bibitem{yellin2002} S.~Yellin, Phys. Rev. D \textbf{66} (2002) 032005, \texttt{arXiv:physics/0203002}.
\bibitem{kims} H.~S.~Lee et~al., PRL \textbf{99} (2007) 091301, \texttt{arXiv:0704.0423}.
\bibitem{edelweiss} V.~Sanglard et~al., Phys.Rev. D \textbf{71} (2005) 122002, \texttt{arXiv:astro-ph/0503265}.
\bibitem{angle2007} J.~Angle et~al., PRL \textbf{100} (2008) 021303, \texttt{arXiv:0706.0039}.
\bibitem{ahmed2008} Z.~Ahmed et~al., \textit{submitted to PRL} (2008), \texttt{arXiv:0802.3530}.
\bibitem{burgess2001} C.~P.~Burgess, M.~Pospelov and T.~ter~Veldhuis, Nucl. Phys. B \textbf{619} (2001) 709-728, \texttt{arXiv:hep-ph/0011335}.
\bibitem{roszkowski2007} L.~Roszkowski, R.~R.~de~Austri and R.~Trotta, JHEP \textbf{07} (2007) 075, \texttt{arXiv:0705.2012}.
\end{thebibliography}
\end{document}